\documentclass[conference]{IEEEtran}
\IEEEoverridecommandlockouts
\usepackage{cite}
\usepackage{amsmath,amssymb,amsfonts}
\usepackage{algorithmic}
\usepackage{graphicx}
\usepackage{textcomp}
\usepackage{xcolor}
\usepackage{orcidlink}
\def\BibTeX{{\rm B\kern-.05em{\sc i\kern-.025em b}\kern-.08em
    T\kern-.1667em\lower.7ex\hbox{E}\kern-.125emX}}

\usepackage{algorithm}
\usepackage{algorithmic}
\usepackage{threeparttable}    
\usepackage{multirow}
\usepackage{tablefootnote}

\usepackage{lipsum}
\usepackage{fancyhdr}
\fancypagestyle{sec}{\lhead{\tmpx}}

\fancypagestyle{arXiv}
{
   \fancyhf{}
   \chead{\textcolor{gray}{This paper will be presented as a lecture at the 2024 IEEE International Symposium on Circuits and Systems (ISCAS), Singapore}}
   \cfoot{ \fontsize{=9pt}{10pt}\selectfont  \textcolor{gray}{© 2024 IEEE. Personal use of this material is permitted. Permission from IEEE must be obtained for all other uses, in any current or future media, including reprinting/republishing this material for advertising or promotional purposes, creating new collective works, for resale or redistribution to servers or lists, or reuse of any copyrighted component of this work in other works}\\\fontsize{=10pt}{12pt}\selectfont\thepage}
   
}

\begin{document}

\title{Epilepsy Seizure Detection and Prediction using an Approximate Spiking Convolutional Transformer

}

\author{
\IEEEauthorblockN{Qinyu Chen\IEEEauthorrefmark{1}\orcidlink{0009-0005-9480-6164}, Congyi Sun \IEEEauthorrefmark{3},
Chang Gao\IEEEauthorrefmark{2}\orcidlink{0000-0002-3284-4078} and
Shih-Chii Liu\IEEEauthorrefmark{1}}
\IEEEauthorblockA{\IEEEauthorrefmark{1}Institute of Neuroinformatics, University of Zurich and ETH Zurich, Switzerland}
\IEEEauthorblockA{\IEEEauthorrefmark{2}Department of Microelectronics, Delft University of Technology, Netherlands}
\IEEEauthorblockA{\IEEEauthorrefmark{3}School of Electronic Science and Technology, Nanjing University, China}
\vspace{-1cm}
\thanks{
 Corresponding authors: Qinyu Chen (q.chen@liacs.leidenuniv.nl) and Chang Gao (chang.gao@tudelft.nl)}
\thanks{
 This work was funded by the Swiss National Science Foundation BRIDGE - Proof of Concept Project (40B1-0\_213731). }
}





\maketitle

\begin{abstract}

Epilepsy is a common disease of the nervous system. Timely prediction of seizures and intervention treatment can significantly reduce the accidental injury of patients and protect the life and health of patients.
This paper presents a neuromorphic Spiking Convolutional Transformer, named Spiking Conformer, to detect and predict epileptic seizure segments from scalped long-term electroencephalogram (EEG) recordings. We report evaluation results from the Spiking Conformer model using the Boston Children’s Hospital-MIT (CHB-MIT) EEG dataset. 
By leveraging spike-based addition operations, the Spiking Conformer significantly reduces the classification computational cost compared to the non-spiking model.
Additionally, we introduce an approximate spiking neuron layer to further reduce spike-triggered neuron updates by nearly 38\% without sacrificing accuracy. 
Using raw EEG data as input, the proposed Spiking Conformer achieved an average sensitivity rate of 94.9\% and a specificity rate of 99.3\% for the seizure detection task, and 96.8\%, 89.5\% for the seizure prediction task, and needs $>$10x fewer operations compared to the non-spiking equivalent model.

\end{abstract}

\begin{IEEEkeywords}
EEG data, epilepsy seizure detection, epilepsy seizure prediction, spiking neural networks, Transformer.
\end{IEEEkeywords}

\section{Introduction}
\thispagestyle{arXiv}
Epilepsy is a disorder of the central nervous system marked by frequent seizures, which are accompanied by abnormal discharge of brain neurons and affect the patient’s behavior.
As indicated by the World Health Organization (WHO), epilepsy is one of the most common long-term neurological conditions that affects more than 50 million people worldwide~\cite{WHO2019}. 
The Electroencephalogram (EEG) is a primary diagnostic tool for clinicians when assessing patients with epilepsy. It is crucial for identifying different phases of EEG signals surrounding a seizure, known as pre-ictal, ictal, post-ictal, and inter-ictal periods. For example, the ictal phase shows sudden, intense electrical bursts corresponding to the seizure event.
Recognizing these EEG signal characteristics is fundamental for developing reliable early detection and seizure prediction systems, enabling more precise treatment, enhanced patient care and disease management


During the past decade, many deep learning techniques have been developed to build seizure detection and prediction systems with high accuracy~\cite{Adam2016Wearable,Muneeb2023SCNN,Omar2020Convolutional,baghdadi2020robust,lammie2021towards, bhattacharya2022epileptic}.
However, these approaches are computationally expensive, making it difficult to implement them on real-world edge devices for point-of-care applications. 
To promote local processing of embedded deep learning models, a variety of model compression techniques have been developed, such as low-precision networks~\cite{zhou2016dorefa} and sparse weights/connectivity patterns~\cite{chen2020efficient,chang2024spartus, chen2022cerebron}. 
A notable alternative is the use of Spiking Neural Networks (SNNs)   
which have emerged as a viable energy-efficient approach if trained properly to produce activation-sparse deep spiking neural networks~\cite{roy2019towards,neil2016learning}.
Recently, several studies on seizure detection have utilized SNN models~\cite{yang2023neuromorphic,shan2023compact,burelo2022neuromorphic}. 
These models use multi-layered MLP or CNN architectures.

This study introduces a novel spiking convolutional transformer, termed the Spiking Conformer, designed to establish a more accurate and efficient model for seizure detection and prediction. It integrates two biologically inspired components: the SNN and the self-attention mechanism. The SNN provides an energy-efficient, event-driven framework, while the self-attention mechanism excels at identifying feature dependencies, enhancing the model's performance.
The proposed Spiking Conformer is trained using raw EEG signals, bypassing noise pre-processing and feature extraction stages, aiming for efficient epileptic seizure detection and prediction.
We benchmarked our model using the CHB-MIT dataset~\cite{chbmit2010} and yielded an average sensitivity of 94.9\% and a specificity rate of 99.3\% for the detection task, and 96.8\%, 89.5\% for the prediction task.
The computational energy efficiency of the algorithm is estimated by calculating the number of operations and shows that our model shows $>$10x reduction in operations compared to the non-spiking convolutional transformer.  

\begin{figure}[tbp]
    \centering
    \includegraphics[width=0.4\textwidth]{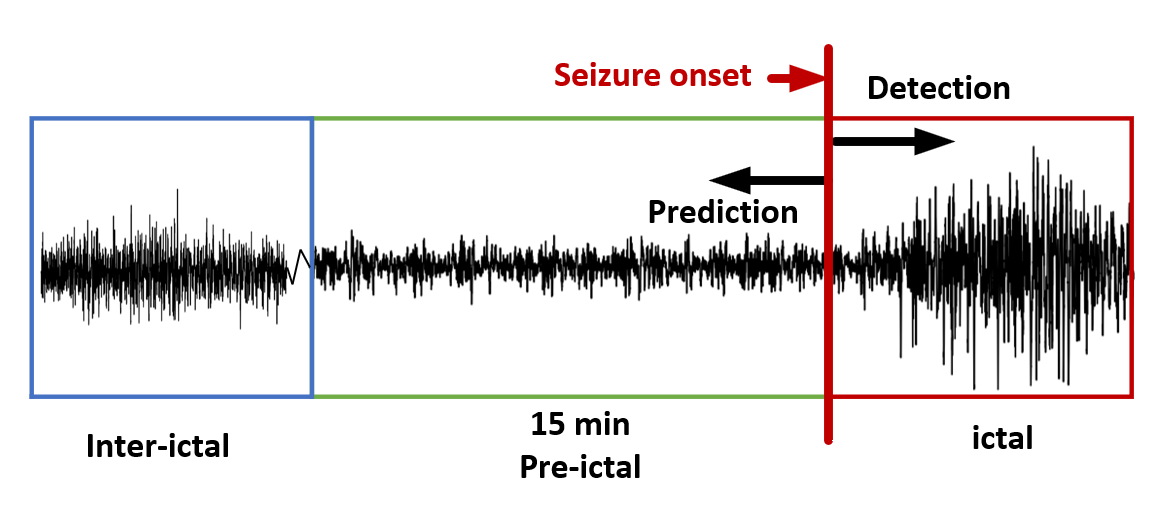}
    \caption{Segment of a patient's scalp EEG from CHB-MIT dataset, showing the different epilepsy states. The figure also illustrates the phases for seizure detection and prediction.
}
    \label{fig:1}
    \vspace{-3mm}
\end{figure}

\begin{figure*}[tbp]
    \centering
    \includegraphics[width=0.75\textwidth]{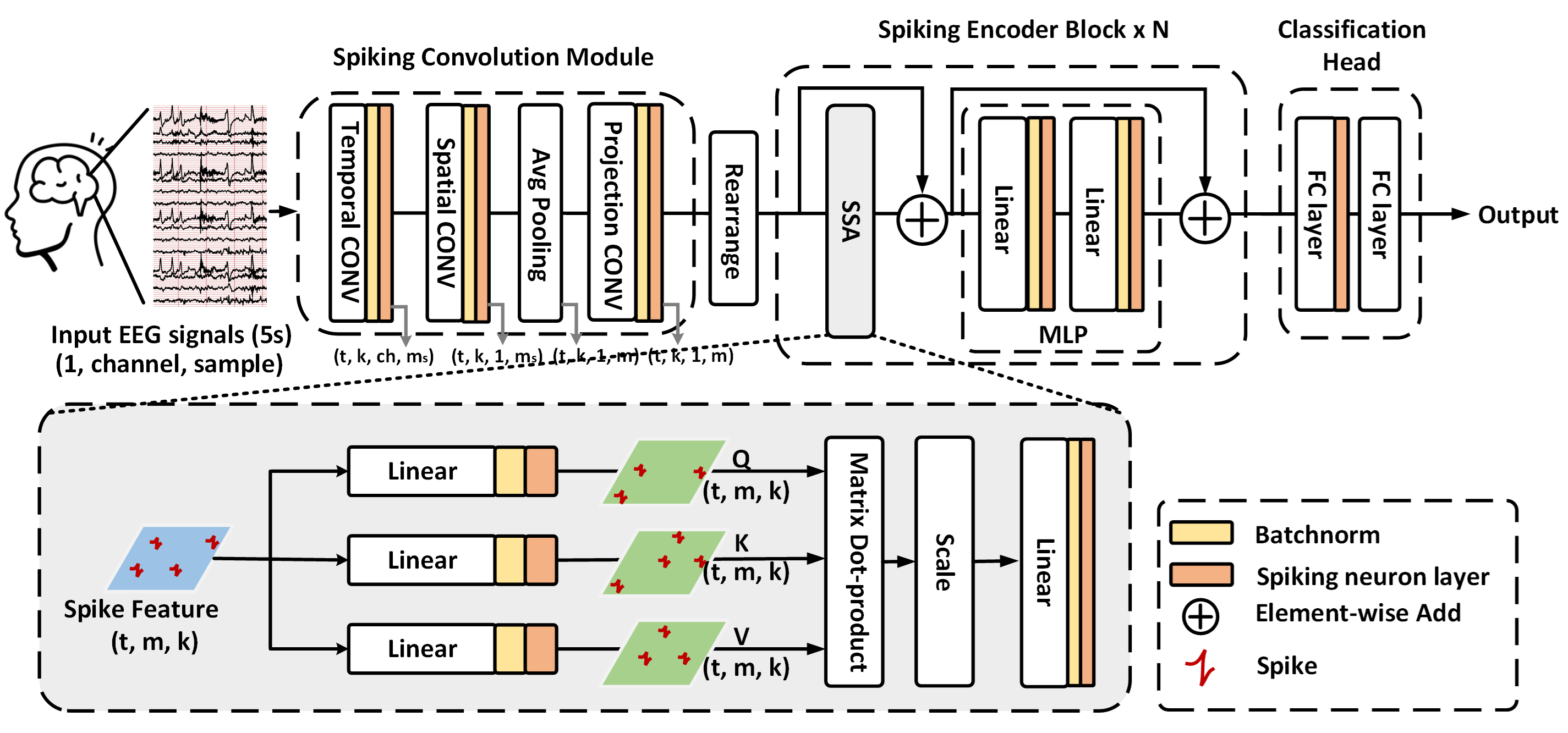}
    \caption{Framework of the proposed Spiking Conformer which includes a spiking convolution module, spiking encoder blocks, and a classification head.
}
    \label{fig:2}
    \vspace{-3mm}
\end{figure*}

\section{METHODOLOGY}
\subsection{Task Description and EEG Dataset }

We evaluate our proposed method on the CHB-MIT scalp long-term EEG dataset which has recordings using 23 scalp electrodes from 23 pediatric patients with intractable seizures at the Boston Children’s Hospital, resulting in 24 cases (with two cases derived from a single patient). Overall, the dataset provides roughly 983 hours of EEG recordings, with individual cases ranging from 19 to 165 hours in duration. 
Expert annotations pinpoint the start and finish of onsets in epochs with ictal activities. All signals are recorded at a sampling frequency of 256 Hz and have a 16-bit resolution.

Figure~\ref{fig:1}  illustrates the distinctions among the pre-ictal, ictal, and inter-ictal phases or states. 
For predicting seizures, the model needs to distinguish between the signals in the pre-ictal and inter-ictal states, and when detecting ongoing seizures, the model should distinguish between the ictal and inter-ictal states. 
It is observed that certain seizures are initiated at the start of the recordings, which hinders our ability to extract extended pre-ictal segments. Given this dataset limitation, we opted for a 15-minute pre-ictal window.
For each recording, we extract those three phases and then apply segmentation in order to obtain 5-second duration segments.

In this dataset, seizures are quite short in duration compared to the full EEG duration in each case, causing an imbalanced data distribution that makes training of the classification network difficult.
To address this imbalance, we increase the number of ictal and pre-ictal segments by using an adjustable stride through long-term signals, thus generating multiple overlapping segments. 
In each case, we aimed for a nearly 1:1 between ictal and pre-ictal to interictal segments to improve the results of the classification. 
After augmenting the data, each phase will yield approximately 30K to 120K segments for each case.
In this work, we use EEG recordings from the first 22 electrodes.


\subsection{Spiking Convolution Transformer for Seizure Detection and Prediction}

Using raw EEG signals eliminates the need for time-intensive feature extraction, making it well-suited for real-time applications. 
In this work, the raw EEG data was presented directly to our network model for end-to-end EEG classification. 
Drawing inspiration from both the CNN and Transformer models, the Spiking Conformer employs spiking convolutions to capture local temporal and spatial characteristics and in addition, utilizes spiking self-attention to capture global temporal feature dependencies. 

An overview of the Spiking Conformer is depicted in Fig.~\ref{fig:2}. 
The architecture comprises three main components: a spiking convolution module, spiking encoder blocks, and a classification head. 
Each input sample is 5\,s long and has a dimension of (1, 22, 1280) corresponding to the number of EEG channels (22) and the sample dimension.

\subsubsection{Spiking Neuron Dynamics}
We use the Leaky Integrate-and-Fire (LIF) spiking neuron model in this work. The discrete updating equations of the LIF model are described as follows:
\begin{equation}
H[t]=V[t-1] + \frac{1}{\tau}(X[t]-(V[t-1]-V_{reset})) 
\label{eq:1}
\end{equation}
\begin{equation}
S[t]=\Theta(H[t]-V_{th})
\label{eq:2}
\end{equation}
\begin{equation}
V[t]=H[t](1-S[t])+V_{reset}\,s[t]
\label{eq:3}
\end{equation}
where $H[t]$ is the membrane potential of the neuron, $s[t]$ is the spiking variable, $\tau$ is the membrane time constant, and $X[t]$ is the input current at time step $t$. When $H[t]$ exceeds the firing threshold $V_{th}$, a spike is triggered, and $S[t]$ is set to 1. $\Theta(V)$ is the Heaviside step function which equals 1 for $V$ $\leq$ 0 and 0 otherwise. $V[t]$ represents the membrane potential after the trigger event which equals $H[t]$ if no spike is generated and otherwise, it is set to the reset potential $V_{reset}$.

\subsubsection{Spiking Convolution Module}
Inspired by~\cite{song2023eegconformer}, we decompose the two-dimensional convolutions into temporal convolutions spanning time intervals and spatial convolutions covering electrode channels.
The first temporal spiking CONV layer uses $k$ kernels, each sized (1, 25) and a stride of (1, 1), applied on the time dimension. 
The subsequent spiking spatial CONV layer also utilizes $k$ kernels, sized ($ch$, 1) with a stride of (1, 1), where $ch$ represents the number of EEG electrode channels. This layer serves as a spatial filter, capturing the interactions among different electrode channels. 
To enhance training and to mitigate overfitting, batch normalization is incorporated. 
An average pooling layer in the time dimension with a kernel size of (1, 64) and a stride of (1, 50) is applied to help enhance signal fidelity and generalization and to reduce computational costs.
A projection spiking CONV layer is then applied to transform the spiking feature maps into the transformer embedding dimensional feature map. 
Finally, we rearrange the feature maps produced by the spiking convolution module, squeeze the electrode channel dimension, and swap the convolution channel dimension with the time dimension. 
This arrangement ensures that the complete feature channels of each temporal point are presented as a token to the spiking encoder module.

\subsubsection{Spiking Encoder Module with Softmax-free Self-attention}

Efforts are being made to explore energy-efficient self-attention \cite{chen2022enabling}.
Similar to the standard transformer encoder block, the spiking encoder block consists of a Spiking self-attention (SSA) block and an MLP block. 
Both the SSA and MLP blocks utilize residual connections.
As the main component in the spiking conformer encoder block, SSA offers an efficient method to model the local-global information of signals using spike-form Query (Q), Key (K), and Value (V) without softmax~\cite{zhou2023spikformer}. 
The input currents of Q, K, and V neurons are computed through learnable matrices first. Subsequently, they are transformed into spiking sequences through spike neuron layers where neuronal charge, fire, and reset are conducted.
The spike-sequence-form Q, K, and V outputs are naturally non-negative (0 or 1), resulting in a non-negative attention map. SSA only aggregates these relevant features and ignores the irrelevant information. Hence it does not need the softmax to ensure the non-negativeness of the attention map.

\subsubsection{Classification Head}
Finally, we use two fully-connected layers as the classifier, which outputs a two-dimensional vector. 
The cross-entropy loss function was utilized to determine the distance between the network output value and the one-hot vector ground truth.
Here, (0,1) represents the inter-ictal phase, while (1,0) signifies the ictal or pre-ictal phase.

\begin{algorithm}[t]
\caption{Approximate Spike-Triggered Neuron Updates}
\label{alg:1}
\begin{algorithmic}[1]
\REQUIRE Total number of timesteps $T$, Threshold timestep $T_{th}$, Weight Matrix $W$, Input current $X$, Spike variable $S$, Spiking Conformer model $model$. 
\FOR{$e$ in $Epochs$}
    \STATE Set $model$ to training mode with $model.train()$.
    \STATE Train $model$ from scratch using the surrogate gradient. 
    \STATE Set $model$ to evaluation mode with $model.eval()$.
    \FOR{$t = 1$ to $T$} 
        \IF{$t \leq T_{th}$}
            \STATE $X[t] \gets W*S[t]$.
            \STATE Determine indices where $X[t] > 0$ and append to $PosIdx$.
        \ELSIF{$t > T_{th}$}
            \STATE Only calculate $X[t][PosIdx]$.
        \ENDIF
    \ENDFOR
\ENDFOR
\end{algorithmic}
\end{algorithm}

\subsection{Approximate Spiking Neuron Layer}
In SNNs, information is encoded and processed using trains of spikes. 
Each neuron is associated with a membrane potential, and a spike is dynamically generated when the potential goes above a specified threshold (see Eqs.~\ref{eq:1}-\ref{eq:3}). 
When a neuron generates a spike, the membrane potential of all its target neurons is incremented by the weights of the respective connections. Thereby, the spike-triggered neuron updates are the most computation-intensive operations.
It is worth noting that we observe that the spiking activity of the SNN model is sparse. In many instances, certain neurons never spike at all. 
Thus the spike-triggered neuron update computations associated with non-spiking neurons can be omitted to achieve faster processing without compromising the accuracy of the network's output.

We develop an approximate methodology, shown in Algorithm~\ref{alg:1}, to identify the potential unnecessary spike-triggered updates and to skip them for reduced computations. 
We train the Spiking Conformer from scratch using the surrogate gradient method described in~\cite{SpikingJelly}.
During the evaluation phase, we track the position, denoted as $PosIdx$, of neurons that fire in the initial $T_{th}$ timesteps, and only update the input currents $X[t]$ of these neurons in the subsequent timesteps. 
This approach is based on the observation that neurons that fire earlier often play a pivotal role in shaping the network's overall response, indicating their importance in handling the most salient features of the input.

\begin{figure*}[tbp]
    \centering
    \includegraphics[width=0.8\textwidth]{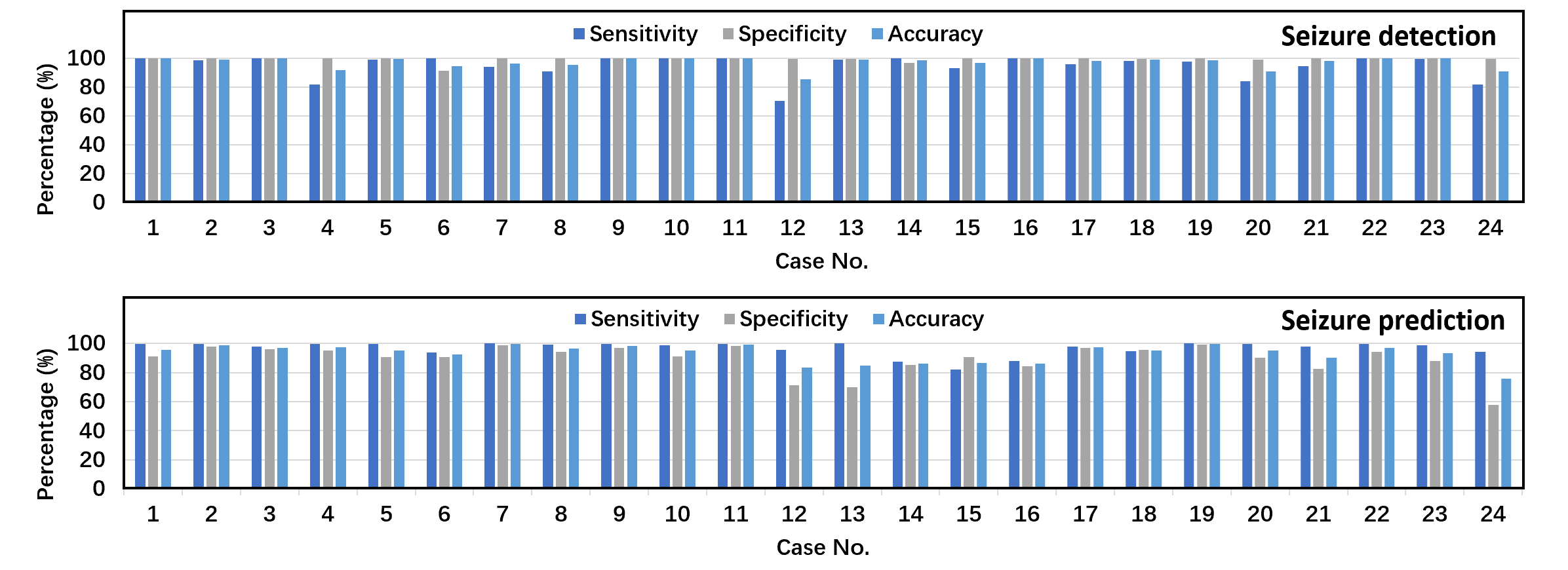}
    \caption{Seizure detection and prediction results on the CHB-MIT EEG dataset (5s window) from the Spiking Conformer.
}
    \label{fig:3}
    \vspace{-3mm}
\end{figure*}

\begin{table*}[t]\renewcommand\arraystretch{0.8}
\caption{Performance Comparison of the Proposed Model with Other Works on Seizure Detection using the CHB-MIT dataset}
\label{tab:1}
\begin{center}
\begin{threeparttable}
\setlength{\tabcolsep}{1.0mm}{
\begin{tabular}{c|c|c|c|c|c|c|ccc}
\hline
\hline
\multirow{2}{*}{\textbf{Ref}} & \multirow{2}{*}{\textbf{Cases}} & \multirow{2}{*}{\textbf{Model}}& \multirow{2}{*}{\textbf{Parameter}}&\multirow{2}{*}{\textbf{Timestep}}& \multirow{2}{*}{\textbf{Feature Extraction}\footnotemark[1]}& \textbf{Validation}&\multicolumn{3}{c}{\textbf{Detection model}}  \\
& & &&&& \textbf{Method}&SENS & SPEC & ACC \\
\hline

Zhou et al. 2018~\cite{zhou2018epileptic}  & 24 &CNN &-&- & FFT & 6-Fold CV  & 96.9 & 98.1 &97.5  \\
Shen et al. 2022 ~\cite{shen2021EEG} & 16 & SVM+RBTE\footnotemark[2]  &-&-& DWT & LOOCV\footnotemark[3]+5-Fold CV & 96.2 & -  & 96.4    \\
Shan et al. 2023~\cite{shan2023compact}  & 24  & SNN+2DLSVM &14.9K & 200 & Power Spectrum & 5-Fold CV & 88.4 & 84.6  & 95.1    \\
Muneeb et al. 2023~\cite{Muneeb2023SCNN}& 24 & Spiking CNN &-& 10 & SE+PLV & - & 95.0 & 99.4 &- \\
\textbf{Our work}  & 24 & Spiking Conformer & 9.9K& 8 & Raw EEG& 10-Fold CV & 94.9 & 99.3 & 97.1  \\
\hline
\end{tabular}}
\begin{tablenotes}
       \footnotesize
       \item[1] DWT denotes Discrete Wavelet Transform, SE denotes Spectral Energy, and PLV denotes Phase Locking Value.
       \item[2] RUSBoosted Tree Ensemble.
       \item[3] Leave-One-Out cross validation.
\end{tablenotes}
\end{threeparttable} 
\end{center}
\end{table*}

\begin{table*}[t]\renewcommand\arraystretch{0.8}
\caption{Performance Comparison of the Proposed Model with Other Works for Seizure Prediction using the CHB-MIT dataset}
\label{tab:2}
\begin{center}
\begin{threeparttable}
\setlength{\tabcolsep}{1.0mm}{
\begin{tabular}{c|c|c|c|c|c|c|ccc}
\hline
\hline
\multirow{2}{*}{\textbf{Ref}} & \multirow{2}{*}{\textbf{Cases}} & \multirow{2}{*}{\textbf{Model}}& \multirow{2}{*}{\textbf{Parameter}}&\multirow{2}{*}{\textbf{Timestep}}& \multirow{2}{*}{\textbf{Feature Extraction}\footnotemark[1]}& \textbf{Validation} &\multicolumn{3}{c}{\textbf{Prediction model}} \\
& & &&& &\textbf{Method}&SENS & SPEC & ACC \\
\hline
Haidar et al. 2018~\cite{Haidar2018Focal}  & 15 & CNN & 185.4K&-& DWT & 10-Fold CV & 87.8 & - & -  \\
Zhang et al. 2020~\cite{zhang2019epilepsy}  & 15 & CNN &194.6K&-& CSP  & LOOCV &  92.2 & 92.0 & 90.0     \\
Buyukccakir et al. 2020~\cite{buyukccakir2020hilbert}  & 10 & MLP & 40.9K&-& HVD  & 10-Fold CV & 89.9 & - & -  \\
Baghdadi et al. 2020~\cite{baghdadi2020robust}&  24 & Deep LSTM & $>$3M &-& Raw EEG & 10-Fold CV  & 84.0 & 90.0 & 88.9\\
Tian et al. 2021~\cite{Tian2021new}& 7 & Spiking CNN  &10.3K &10& Raw EEG &80-20 split& 95.1 & 99.2 & -  \\
Zhao et al. 2022~\cite{Zhao2022patient}  & 19 & AdderNet-SCL & 120K&-& Raw EEG & LOOCV & 94.9 & - & -  \\
Lu et al. 2023~\cite{Lu2023Epileptic}& 11 & CBAM-3D CNN-LSTM  &- &-& STFT &LOOCV& 98.4 & - & 97.9  \\
\textbf{Our work}  & 24 & Spiking Conformer & 40.3K &8& Raw EEG& 10-Fold CV  & 96.8   & 89.5  & 93.1 \\
\hline
\end{tabular}}
\begin{tablenotes} 
\item[1] CSP denotes Common spatial pattern, STFT denotes Short-Time Fourier Transform, HVD denotes Hilbert Vibration Decomposition.
\end{tablenotes}
\end{threeparttable} 
\end{center}
\end{table*}

\section{Experimental Results}
We evaluated the Spiking Conformer model on the CHB-MIT database using the raw EEG data and applied a 10-fold cross-validation technique by randomly shuffling the trials and partitioning them into 10 unique subsets. In each iteration, nine of these subsets were combined to form the training set, while the remaining subset was used as the test set. To evaluate and compare the results of our network with other state-of-the-art methods, we utilize the following metrics:
sensitivity (SENS), specificity (SPEC), and accuracy (ACC).
SENS measures the model's ability to correctly identify actual pre-ictal (ictal) segments, higher sensitivity means fewer missed pre-ictal (ictal) cases.
SPEC indicates how well the model identifies actual inter-ictal segments, higher specificity reflects fewer false alarms for pre-ictal (ictal) states.

Figure~\ref{fig:3} shows the SENS, SPEC, and ACC metrics across all 24 cases from the CHB-MIT dataset. 
For the detection task, we employ a 9.9K parameter 8-timestep Spiking Conformer model, which contains only one spiking encoder and $k= 8$ in the spiking convolution module. The average SENS, SPEC, and ACC values are 94.9\%, 99.3\%, and 97.1\%, respectively. 
For the prediction task, we employ a 40.3K parameter 8-timestep Spiking Conformer model, which contains two spiking encoders and $k=32$ in the spiking convolution module. The average SENS, SPEC, and ACC values are 96.8\%, 89.5\%, and 93.1\%, respectively. 
In the proposed approximate spiking neuron layer, we set $T_{th}=2$ and find that 
the number of spike-triggered neuron update operations was reduced by 37.9\% and 35.9\% for each task respectively, without any performance metric loss.

We also evaluated the computational efficiency by comparing the number of operations in our Spiking Conformer with a non-spiking convolutional transformer. For the prediction task, the Spiking Conformer requires 2.1M ADD and 6.1K MUL operations, while the conventional transformer needs a total of 27.1M operations, combining both MUL and ADD. In the detection task, the Spiking Conformer requires 0.32M ADD and 1.0K MUL operations, in contrast to the conventional transformer's 4.1M combined MUL and ADD operations.

Tables~\ref{tab:1} and~\ref{tab:2} compare our model with recent state-of-the-art works for the seizure tasks using the same dataset. 
For seizure detection, our Spiking Conformer with raw EEG data input and fewer timesteps, delivered comparable SENS and SPEC compared to the spiking CNN~\cite{Muneeb2023SCNN}, even though the latter employed an extra feature extraction process.
For seizure prediction, our spiking model achieved the highest sensitivity among those models \cite{baghdadi2020robust,Tian2021new, Zhao2022patient} using raw EEG signals.

\section{Conclusion}
This work introduces the Spiking Conformer tested on the CHB-MIT database. 
The approximate spiking neuron layer helps reduce computational operations by nearly 38\%
in both tasks without sacrificing accuracy. Compared to the equivalent non-spiking convolution transformer, the Spiking Conformer requires $>$10x fewer operations. This showcases the Spiking Conformer's potential in efficiently advancing epilepsy diagnostics on embedded systems.

\bibliographystyle{IEEEtran}
\bibliography{refs.bib}
\vspace{12pt}

\end{document}